\documentclass{elsart}
\usepackage{epsf}     

\begin{document}
\begin{frontmatter}

\title{
Currents Induced by Charges Moving in Semiconductor 
}

\author{I.V.Kotov\thanksref{CA}}
\address{The Ohio State University, Columbus, OH 43210, USA}
\thanks[CA]{ Phone: (614)--292--4775; fax: (614)--292--4833;
e--mail: kotov@mps.ohio-state.edu; 
This work was supported in part by 
NSF grant PHY--0099476.
}

\end{frontmatter}

\section*{Introduction}

The method of computation of currents induced on electrodes 
by charges moving in vacuum was introduced in \cite{Shockley}, \cite{Ramo}.
In presented paper, this method is generalised for charges moving in a media.

\section*{Derivation of the Induced Current}

Consider the moving charge, $q_m$, inside a semiconductor
and any number of conductors at constant potentials
for one of which, say A (following Ramo's approach \cite{Ramo}),
the induced current is desired.
The potential of the electrostatic field inside the semiconductor 
under static or low frequency conditions \cite{Sze}
satisfies

\begin{equation}
\varepsilon {\varepsilon}_0 {\nabla}^{2}V = - \rho , \label{eq:LaM}
\end{equation}

%

where $\rho$ is a charge density, 
${\varepsilon}_0$ is permittivity of free space and
$\varepsilon$ is permittivity of the media.

The boundaries are surfaces of electrodes and
tiny equipotential sphere surrounding moving charge.
The integral over a boundary surface $S_i$ equals to

\begin{equation}
	-\int_{S_i}^{} 
	{ \varepsilon {\varepsilon}_0 {\frac{\partial V }{\partial n}} ds } 
	= Q_i , 
\label{eq:Gs}
\end{equation}

where $n$ is the outward normal to the surface
and $Q_i$ is the charge on the i-th electrode
or moving charge.

Also consider the same geometry of conductors in the free space
but without moving charge.
Potential of the electrode A is rised to unit potential
and all other conductors are grounded.
Potential in this case, $V'$, satisfies
${\nabla}^{2}V' = 0$ everywhere between conductors.

Applying Green's theorem 
%
%
to the vector function

\begin{equation}
{\bf f} =  \varepsilon {\varepsilon}_0 
( V' \cdot {{\nabla}V} - V \cdot {{\nabla}V'} ),  \label{eq:fg}
\end{equation}

we obtain the following equation

\begin{equation}
	\int_{Volume}^{} 
	{ \varepsilon {\varepsilon}_0 ( V' \cdot {{\nabla}^2 V} - V \cdot {{\nabla}^2 V'} ) dv} 
	=
	- \int_{Surface}^{} 
	{ \varepsilon {\varepsilon}_0 
	  ( V' {\frac{\partial V }{\partial n}} - 
	   V  {\frac{\partial V'}{\partial n}}) ds }  
\label{eq:Gt}
\end{equation}

Using, that by design, $V' = 0$ on the surfaces of
all electrodes but A and 
${\nabla}^{2}V' = 0$ in the space between boundaries
Eq.~\ref{eq:Gt} can be rewritten as

\begin{equation}
	\int_{Volume}^{} 
	{ \varepsilon {\varepsilon}_0  V' \cdot {\nabla}^2 V \cdot  dv} 
	=
	Q_A + V'_m \cdot q_m  - 
	\sum_{electrodes}^{}   
	{ \varepsilon V_i Q'_i }  
\label{eq:Gm}
\end{equation}

or

\begin{equation}
	Q_A = - q_m \cdot V'_m 
		+ \sum_{electrodes}^{}{ \varepsilon V_i \cdot Q'_i} 
		+ \int_{Volume}^{} 
		{ V' \cdot \varepsilon {\varepsilon}_0 {\nabla}^2 V  \cdot dv} 
\label{eq:Qa}
\end{equation}

\begin{equation}
	i_A =  \frac{dQ_a}{dt} =
		- q_m \cdot \frac{dV'_m}{dt} 
		+ \frac{\partial}{\partial t}
		\int_{Volume}^{} 
		{ V' \cdot \varepsilon {\varepsilon}_0 {\nabla}^2 V \cdot dv} 
\label{eq:ia}
\end{equation}

This can be revritten as

\begin{equation}
	i_A = - q_m \cdot \frac{dV'_m}{d{\bf r}} \frac{d{\bf r}}{dt} 
		+ \int_{Volume}^{} 
		{ V' \cdot \frac{\partial \rho}{\partial t} dv} 
\label{eq:ia_c}
\end{equation}

where $\bf r$ is the direction of the motion.
In terms of weighting field, ${\bf E'} = - \nabla V'$
this became

\begin{equation}
	i_A =  q_m \cdot {\bf v \cdot E'} 
		+ \int_{Volume}^{} 
		{ V' \cdot \frac{\partial \rho}{\partial t} dv} 
\label{eq:ia_f}
\end{equation}

where the integration is over the volume of the media.
In the case when media is vacuum or isolator, the second term is zero  
and Eq.~\ref{eq:ia_f} became Ramo-Shockley's equation. 
In the case of semiconductor, this term is a weighted sum of
currents generated by moving charge inside the semiconductor.

\end{document}